\documentclass[10pt]{article}

\newcommand{\ket}[1]{\big| #1 \big\rangle}  
\newcommand{\bra}[1]{\big\langle #1 \big|}  
\newcommand{\braket}[2]{\big\langle #1 \big| #2 \big\rangle}                 

\usepackage{authblk}
\usepackage{soul}

\usepackage{cite}
\usepackage{amsmath,amssymb,amsfonts}
\usepackage{algorithmic}
\usepackage{graphicx}
\usepackage{textcomp}
\usepackage{xcolor}

\usepackage{physics}

\def\BibTeX{{\rm B\kern-.05em{\sc i\kern-.025em b}\kern-.08em
    T\kern-.1667em\lower.7ex\hbox{E}\kern-.125emX}}
    
\begin{document}

\title{Hiperwalk: Simulation of Quantum Walks with Heterogeneous High-Performance Computing}

\author[1]{Paulo Motta} 
\author[1]{Gustavo A. Bezerra}
\author[2]{Anderson F. P. Santos} 
\author[1]{Renato Portugal} 

\affil[1]{\small{National Laboratory of Scientific Computing,
          Petr\'opolis, RJ, 25651-075, Brazil}
}

\affil[2]{\small{Military Institute of Engineering,
Rio de Janeiro, RJ, 22290-270, Brazil}
}

\maketitle

\begin{abstract}
The Hiperwalk package is designed to facilitate the simulation of quantum walks using heterogeneous high-performance computing, taking advantage of the parallel processing power of diverse processors such as CPUs, GPUs, and acceleration cards. This package enables the simulation of both the continuous-time and discrete-time quantum walk models, effectively modeling the behavior of quantum systems on large graphs. Hiperwalk features a user-friendly Python package frontend with comprehensive documentation, as well as a high-performance C-based inner core that leverages parallel computing for efficient linear algebra calculations. This versatile tool empowers researchers to better understand quantum walk behavior, optimize implementation, and explore a wide range of potential applications, including spatial search algorithms.

\

\noindent
Keywords: quantum walk, simulation, HPC, graphs
\end{abstract}

\section{Introduction}

This paper\footnote{ https://doi.org/10.1109/QCE57702.2023.00055 -- Quantum Week 2023} introduces a new and updated version of Hiperwalk\footnote{\textbf{https://hiperwalk.org}}, a freeware open-source program designed for simulating quantum walks on graphs using high-performance computing (HPC) on classical computers. Quantum walks provide an excellent framework for exploring quantum information processing~\cite{NC00}, allowing the development of innovative quantum algorithms and enhancing our understanding of quantum dynamics in complex systems. Compared to the previous version~\cite{LLP15}, which used separated processes to execute linear algebra computations using the Neblina\cite{neblina} programming language, the new version is a complete overhaul. It enables users to simulate both the coined quantum walk~\cite{AAKV01} and continuous-time quantum walk~\cite{FG98} models on graphs in a Python session. Additionally, its ability to utilize all the parallel devices on a user's machine makes it a versatile tool for researchers in the fields of quantum walks and quantum algorithms.

Hiperwalk consists of two primary components: a user frontend, which serves as an application programming interface (API), and an inner core, which leverages the original Neblina\cite{neblina} code by turning it into a modular library. The user frontend API is developed in Python, offering an accessible and user-friendly interface that accepts the adjacency matrix of a graph as input for quantum walk applications. This frontend API then prepares the system for simulating the quantum walk dynamics, ensuring that the necessary data structures and initial conditions are properly established. Additional Python commands can be employed to supplement the built-in functionalities provided by the Hiperwalk package.

The inner core of Hiperwalk is written in C and leverages HPC for efficient calculations. Upon receiving the input from the frontend, the core performs all linear algebra calculations in parallel, using the devices available on the user's machine. This parallel processing approach enables Hiperwalk to handle large-scale simulations and complex graphs with ease, significantly reducing the computational time required for simulating quantum walk dynamics.

This paper provides a detailed description of Hiperwalk's architecture, including the implementation of both the coined and continuous-time quantum walk models. We also present benchmark results that showcase the efficiency and scalability of the package across various graph sizes and topologies. Moreover, we explore the potential applications of Hiperwalk, such as the development of quantum algorithms and spatial quantum search~\cite{Por18book}.

In summary, Hiperwalk is a powerful and versatile tool for simulating quantum walks on graphs, offering researchers an efficient and accessible way to explore the potential of quantum walks in analyzing complex systems. Users can utilize all available parallel devices on their machines, including CPUs and graphics cards. The open-source nature of the package fosters collaboration and innovation within the community, encouraging the development of new features and enhancements. We invite researchers to take advantage of Hiperwalk and contribute to its growth, driving new discoveries and advancements in the rapidly evolving domain of quantum computing.

In this paper, we have organized our discussion as follows:
Section~\ref{sec:2} presents a review of the literature relevant to our research, including previous studies and approaches in the field. We will discuss how our work builds upon or differs from these existing works.
Section~\ref{sec:3} describes how to use Hiperwalk to simulate the time evolution of continuous-time quantum walks and coined quantum walks. We provide examples to illustrate the process and showcase the functionality of our tool.
Section~\ref{sec:4} delves into the details of the inner core, discussing its design, implementation, and features. We explain the improvements made to the core, including the incorporation of new features aimed at enhancing performance, flexibility, and capabilities.
In section~\ref{sec:5}, we discuss the implications of our work, and suggest potential future directions for further investigation and development.

\section{Related Works}\label{sec:2}

The study of quantum walks on graphs has aroused the interest of developers, leading to the creation of several packages that offer a wide range of features for calculating and visualizing the time evolution of quantum walks. Researchers in this area can take advantage of these advanced tools to effectively examine the behavior and properties of quantum walk models, enhancing our understanding and application of quantum walks across various fields. Important packages in this domain, each built to a specific context of the quantum walk dynamics, include QWalk~\cite{MP08}, QwViz~\cite{BBW11}, pyCTQW~\cite{IW15}, and QSWalk~\cite{FRW17}.

Since most of those packages do not leverage HPC explicitly, we find related works when we look at the area of high-performance simulation of quantum computing~\cite{10.1007/978-3-319-27119-4_17,DERAEDT2007121,DERAEDT201947,Jones2019} and of quantum circuits~\cite{CHEN2018964,Villalonga2019,Guerreschi2020}. Note that those references lack commands for quantum walks.

Our ongoing work focuses on revitalizing the Hiperwalk simulator. To achieve this, we need to modify the way we compute quantum operations and supply the high-performance linear algebra resources necessary for executing the computations efficiently.

As discussed in~\cite{10.1145/2661136.2661156, UCAM-CL-TR-902, 1438333}, we can generally conclude that providing a library is more favorable than creating a new programming language. With this in mind, we can consider using existing libraries like~\cite{10.1145/3204919.3204924}, which offer a Basic Linear Algebra Subprograms (BLAS) implementation based on OpenCL. However, incorporating this would necessitate numerous changes to the Hiperwalk code, as it does not directly interact with low-level code.

Another alternative would be to use a library that directly accesses the GPU from Python but lacks linear algebra support, such as PyCuda/PyOpenCL~\cite{Klockner:2012:PPS:2109228.2109321}. Adopting this type of library would necessitate porting the Neblina code to a combination of Python and embedded C code to create the parallel kernels essential for the computations.

Nonetheless, both solutions have limitations: the first lacks a solution for sparse matrix operations, while the latter necessitates reimplementing linear algebra routines from scratch. Bearing this in mind, we developed a layered model that provides both sparse and dense matrix operations via a unified abstraction API.


Another noteworthy reference is the distribution model adopted in~\cite{10.1145/3419111.3421287}, which leverages the serverless model for distributed computing. Although they employ a matrix tiling strategy for distributed computing, they do not use accelerators and rely solely on CPU power for their computations. Limiting the implementation to CPU-only presents a drawback for us, as our original environment already supported the use of GPUs. Conversely, the data tiling approach is a technique we anticipate as a viable means to increase the problem size for computation, both locally to overcome GPU memory constraints and in a distributed environment.

As discussed throughout the text, numerous approaches and libraries use GPUs to accelerate the computation of linear algebra functions. However, most either concentrate on solving a specific problem efficiently or offer a more generic solution that does not support sparse matrices. We believe our contribution is crucial in bridging this gap, as we unify both worlds by employing different linear algebra implementations and making them accessible to users through the same API our library provides.

As a final remark, it is important to mention that our approach uses sparse matrices to allow computation of large-size quantum walk problems, which, in turn, makes a comparison with quantum processor units, at most, speculative and non-productive. However, we believe that such a comparison will make sense in the future.

\section{User Frontend}\label{sec:3}

Quantum walks are an important area of research in quantum computing and have applications in various fields, such as quantum algorithms, quantum search, and simulation of complex systems. Hiperwalk is a package that enables users to study and simulate the time evolution of continuous-time quantum walks and discrete-time coined quantum walks using heterogeneous high-performance computing. This section describes the user frontend, which is a Python package, making Hiperwalk highly accessible and easy to use for researchers and graduate students in the field.


Hiperwalk follows the same structure as standard Python packages and provides comprehensive documentation, in line with the methods used by Python libraries. The documentation includes a tutorial outlining the first steps to using the Hiperwalk package, as well as instructions on how to install it. Users can quickly get started with the package, making it a valuable resource for those studying quantum walks.


The Hiperwalk library comprises various Python classes relevant to users. It features two primary classes, each dedicated to simulating the dynamics of continuous-time quantum walks and coined quantum walks. Users can explore and experiment with diverse quantum walks, gaining insights into their unique properties and behaviors.


Among the classes available in Hiperwalk are well-known graph structures such as cycles, two-dimensional grids, hypercube graphs, and more. Additionally, a generic \textit{Graph} class allows users to simulate time evolution on arbitrary graphs by inputting the adjacency matrix of the graph. This flexibility enables researchers to study a wide range of graph structures and their impact on quantum walk dynamics.


Hiperwalk provides functions to display the probability distribution of quantum walks, offering valuable insights into their behavior. Furthermore, the package includes methods to animate the time-evolution of quantum walks, allowing users to observe the dynamics in real-time and develop a deeper understanding of the process.


The Hiperwalk package also includes methods for implementing quantum walks in the context of spatial search algorithms.

\subsection{Continuous-time quantum walks}

The dynamic of the continuous-time quantum walk on a graph with adjacency matrix $A$ is driven by a Hamiltonian $H$ given by
\begin{equation*}
H = -\gamma A - \sum_{v\in M}\ket{v}\bra{v},
\end{equation*}
where $\gamma$ is a positive parameter and $M$ is the set of marked vertices ($M$ can be the empty set)~\cite{CG04}.  The time-dependent evolution operator is given by
$$U(t)=\text{e}^{-\text{i}Ht},$$
and the state of the walk at time $t$ is $\ket{\psi(t)}=U(t)\ket{\psi(0)}.$ The probability of finding the walker on vertex $v$ at time $t$ is $\left|\braket{v}{\psi(t)}\right|^2$.

This dynamic is implemented in Hiperwalk using the class \texttt{ContinuousTime}, which has methods that assist in analyzing the time evolution of the continuous-time quantum walk. An instance of this class represents a continuous-time quantum walk on a specific graph (characterized by the adjacency matrix), a positive parameter $\gamma$, and an optional list of marked vertices.

To create an instance named \texttt{ctqw} of a continuous-time quantum walk on a graph with adjacency matrix $A$, $\gamma=0.35$, and marked vertices with labels 1 and 4, execute the following command:

\vspace{4.6pt}
\noindent
[1]:~$\texttt{ctqw = hiperwalk.ContinuousTime(} \texttt{graph=A, gamma=0.35, time=0.03,}\texttt{ marked=[1,4])}$
\vspace{4.6pt}

\noindent
An adjacency matrix $A$ can be created using the \texttt{NetworkX}\footnote{https://networkx.org/} package or alternatively through direct Python commands, with the optional support of the \texttt{NumPy}\footnote{https://numpy.org/} or \texttt{SciPy}\footnote{https://scipy.org/} libraries. For this example, we assume that the graph has at least five vertices ($A$ must have a minimum of five rows). If the graph has no marked vertices, the third argument can be omitted.

If the user wants to view the Hamiltonian of the walk explicitly, it can be obtained by executing the following command:

\vspace{4.6pt}
\noindent
[2]:~\texttt{H = ctqw.get\_hamiltonian()}
\vspace{4.6pt}

%

To simulate the time evolution, we need an initial condition, which is a state (normalized $n$-dimensional vector, where $n$ is the number of vertices). There is a method called \textit{ket} that aids in creating a state using vectors of the computational basis. For instance, the commands

\vspace{4.6pt}
\noindent
[3]:~\texttt{a = 1/2$^{**}$0.5}

\noindent
[4]:~{\texttt{psi0 = a$^*$(ctqw.ket(2)+1j$^*$ctqw.ket(4))}
\vspace{4.6pt}

\noindent
create a normalized superposition of vertices 2 and 4
with amplitudes $1/\sqrt 2$ and $i/\sqrt 2$, respectively.
Now, we are ready for the simulation. The command

\vspace{4.6pt}
\noindent
[5]:~\texttt{states = ctqw.simulate(}$\texttt{range=21,}$ 
 $\texttt{state=psi0)}$
\vspace{4.6pt}

\noindent
calculates the evolution operator using the Hamiltonian, and using the evolution operator
calculates the states $\ket{\psi(k\,\Delta t)}$ for $k=0,1,..., 20$ taking $\Delta t=0.03$. The output is the list of the states.

We calculate the list of probability distributions by executing the command

\vspace{4.6pt}
\noindent
[6]:~\texttt{probs = ctqw.probability\_distribution(}$\texttt{states)}$
\vspace{4.6pt}

\noindent
The last command returns a list of probability distributions. The $i$-th probability distribution in \texttt{probs} corresponds to the $i$-th state in \texttt{states}. The probability distribution for a continuous-time quantum walk state can be determined by taking the square of the absolute value of each individual entry.


The output of the last command can be used for
plotting the probability distributions individually or
generating an animation.
A Hiperwalk function is available for performing this task.
Executing the command

\vspace{4.6pt}
\noindent
[7]:~\texttt{hiperwalk.plot\_probability\_distribution(}$\texttt{probs, animate=True)}$
\vspace{4.6pt}

\noindent
outputs an animation.
This command accepts some optional arguments from the \texttt{Matplotlib}\footnote{https://matplotlib.org/} and \texttt{NetworkX} libraries for plotting customization.

\begin{figure}[!ht]
    \centering
\includegraphics[scale=0.22]{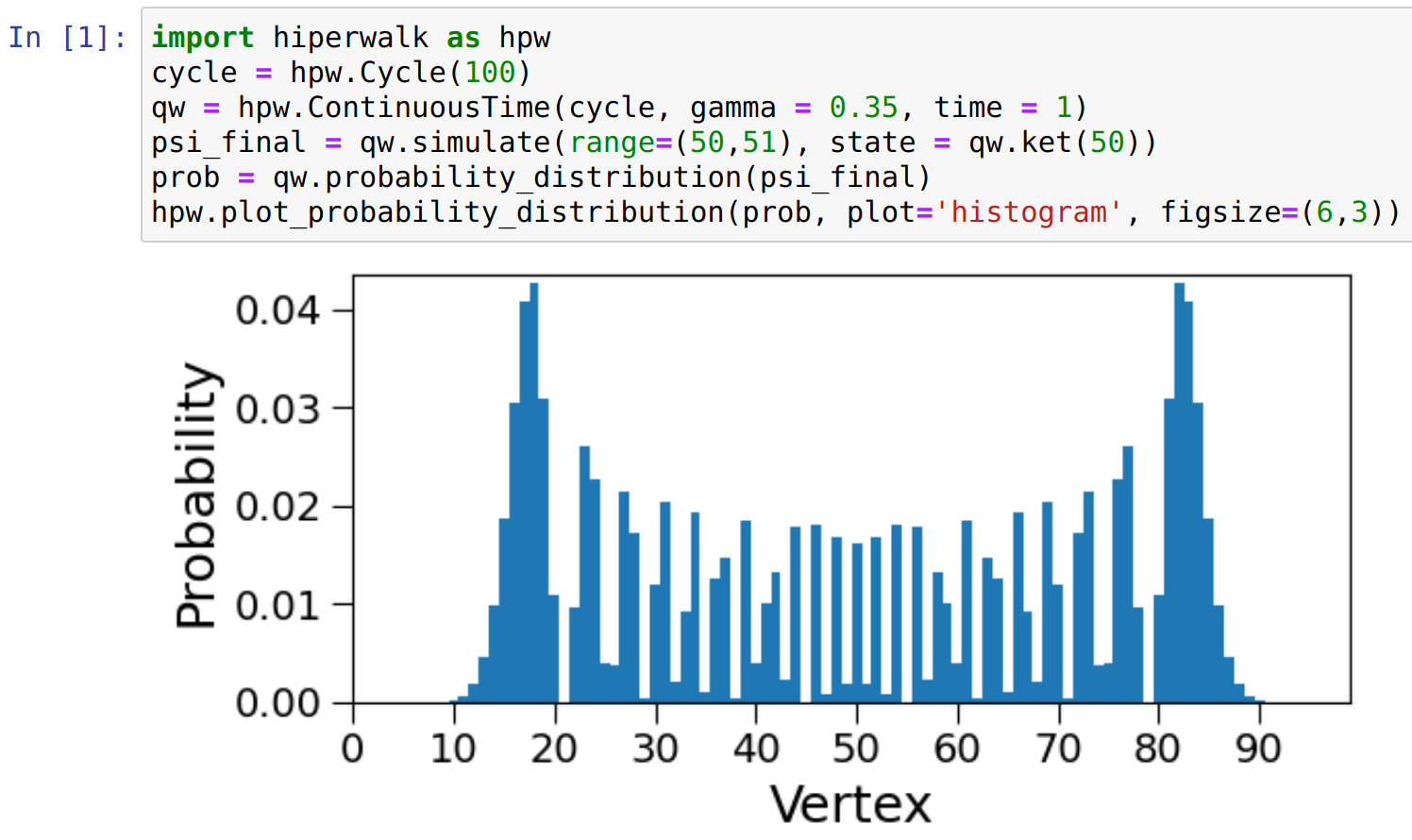}
    \caption{Excerpt from a Jupyter Notebook showing the commands that plot the probability distribution of a continuous-time quantum walk at $t=50$.}
    \label{fig:fig1}
\end{figure}

We have described the \texttt{ContinuousTime} class, which allows users to analyze the time evolution of continuous-time quantum walks on arbitrary graphs. The library includes numerous \textit{classes} for specific graph types, such as \texttt{Cycle}, \texttt{Line}, \texttt{Grid}, \texttt{Hypercube}, and more.
Figure~\ref{fig:fig1} depicts an excerpt from a Jupyter Notebook showing the probability distribution of a continuous-time quantum walk on a cycle with 101 vertices at time $t=50$
with the walker departing from vertex 50. Note that we used a \texttt{Cycle} class instance instead of the graph adjacency matrix. This can be done for any specific graph class in the library.

The evolution operator is computed using resources provided by the inner core. If the user's machine has a GPU available, then high-performance computing (HPC) will be employed. If not, computations will depend on CPU-based parallel execution.

\subsection{Coined quantum walks}

The dynamic of coined quantum walks is driven by an evolution operator 
$$U=SC,$$
where $S$ is usually the flip-flop shift operator defined as
$$S\ket{v,w}=\ket{w,v},$$ 
where $(v,w)$ is an arbitrary arc with tail $v$ and head $w$, and $C$ is the coin. The Grover coin is defined as
\begin{equation*}
	C\ket{v,w}= \! \sum_{\substack{v'\in{N(v)}}} \!\!\!\! \left(\frac{2}{d(v)}-\delta_{w,v'}\right)\!\ket{v,v'},
\end{equation*}
where $N(v)$ is the set of neighbors of $v$, and $d(v)$ is the degree of $v$. If the initial state is $\ket{\psi(0)}$, the state of the walk after $t$ steps is $\ket{\psi(t)}=U^t\ket{\psi(0)}$. The probability of finding the walker on a vertex $v$ after $t$ steps is
\begin{equation*}
	p_v(t)= \!\sum_{w\in N(v)} \! \left|\braket{v,w}{\psi(t)}\right|^2.
\end{equation*}

This dynamic is implemented using the class \texttt{Coined}, which has methods that assist in analyzing the time evolution of coined quantum walks. An instance of this class represents a discrete-time coined quantum walk on a specific graph, characterized by its adjacency matrix. The implementation of this class adheres to the same structure as the \texttt{ContinuousTime} class whenever possible. 

To create an instance named \texttt{dtqw}
of a coined quantum walk on a graph with adjacency matrix $A$, issue the command

\vspace{4.6pt}
\noindent
[1]:~\texttt{dtqw = hiperwalk.Coined$(A,$} $\texttt{shift='flipflop',}$ $\texttt{coin='grover')}$
\vspace{4.6pt}

\noindent
An adjacency matrix $A$ can be created using the \texttt{NetworkX} package, or alternatively, through Python commands with the optional support of the \texttt{NumPy} or \texttt{SciPy} libraries. Once $A$ is defined, the computational basis consists of a list of arcs ordered as follows: The first arcs have tails labeled 0, i.e., $(0,v)$ for all $v \in N(0)$, where $N(0)$ is the neighborhood of vertex 0 (label 0 corresponds to the first row of the adjacency matrix). These arcs are ordered with respect to the head, i.e., if $u, v \in N(0)$ and $u < v$, then the arc $(0, u)$ precedes $(0, v)$ in the computational basis. The subsequent arcs have tails labeled 1, and so on. The size of the computational basis is $2|E|$, where $E$ denotes the edge set. With this ordering convention, the shift operator is not block-diagonal in general, but the coin operator is, with the dimension of each block being the degree of each vertex. The flip-flop shift operator can be applied to any graph, but in specific graph classes such as \texttt{Line} or \texttt{Grid}, persistent shift operators can be chosen.

After creating an instance \texttt{dtqw} of the \texttt{Coined} class, we can proceed with generating the time evolution. But first, let's define an initial state called \texttt{psi0}. Suppose vertex 0 is adjacent to vertices 1 and 2, and we want the initial state to be a uniform superposition of these arcs. Then,

\vspace{4.6pt}
\noindent
[2]:~\texttt{a = 1/2$^{**}$0.5}

\noindent
[3]:~\texttt{psi0 = a$^*$dtqw.ket(0,1)+a$^*$dtqw.ket(0,2)}
\vspace{4.6pt}

\noindent
are the commands that generate the desired state
$$\ket{\psi(0)}=\frac{\ket{0,1}+\ket{0,2}}{\sqrt 2}.$$

With this in place, we are now prepared for the simulation. The command

\vspace{4.6pt}
\noindent
[4]:~\texttt{states = dtqw.simulate(range=(0,50,5),} $\texttt{state=psi0)}$
\vspace{4.6pt}

\noindent
calculates the evolution operator and
creates a list of states $\ket{\psi(k\,\Delta t)}$ for $k=0,1,...,49$, where $\Delta t=5$. 

We calculate the list of probability distributions by executing the command

\vspace{4.6pt}
\noindent
[5]:~\texttt{probs = dtqw.probability\_distribution(}$\texttt{states)}$
\vspace{4.6pt}

\noindent
The output can be used for generating an animation by executing the command

\vspace{4.6pt}
\noindent
[6]:~\texttt{hiperwalk.plot\_probability\_distribution(}$\texttt{probs, animate=True)}$
\vspace{4.6pt}

\begin{figure}[h!]
    \centering
\includegraphics[scale=0.23]{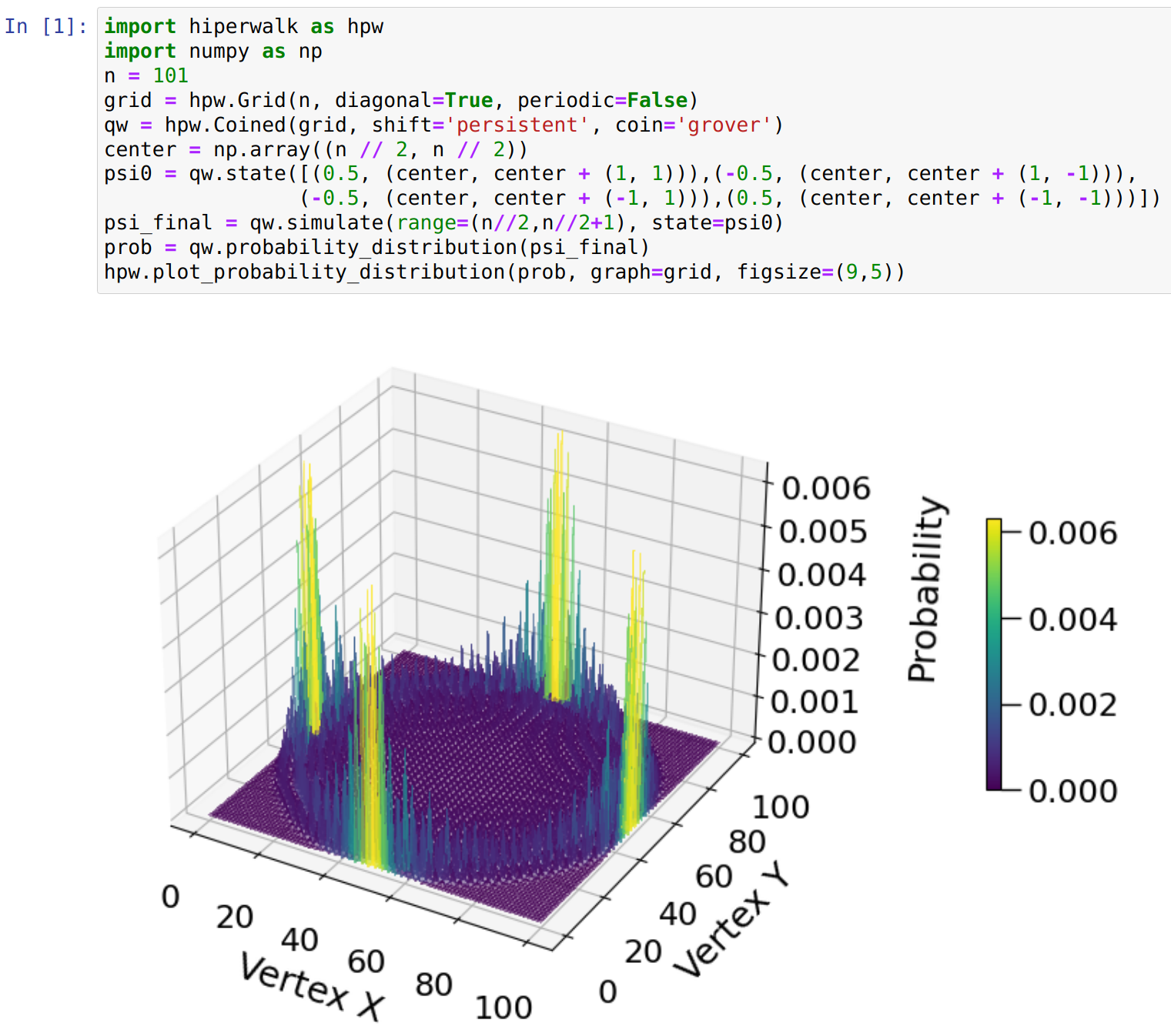}
    \caption{Excerpt from a Jupyter Notebook showing the commands that plot the probability distribution of a coined quantum walk on a two-dimensional grid.}
    \label{fig:fig2}
\end{figure}

Figure~\ref{fig:fig2} shows an application of the package that creates the plot of the probability distribution of a coined quantum walk on a two-dimensional grid. The shift operator is persistent, and the coin is the Grover operator. The walker departs from the center of the lattice depicted in the figure and takes 60 steps. We passed a specific class graph object instead of the adjacency matrix for creating the \texttt{Coined} instance and for the plotting algorithm. This allows the creation of unique operators (e.g., persistent shift), the usage of position-coin notation instead of arc notation and plotting the graph with its specific visualization.


\subsection{Closing remarks}

There are many methods to deal with quantum-walk-based search algorithms on graphs with marked vertices.

In conclusion, the frontend of Hiperwalk is a powerful and user-friendly Python package that caters to the needs of researchers and graduate students working on quantum walks. With its well-documented methods and an array of features for simulating various graph structures and quantum walks in spatial search algorithms, Hiperwalk has the potential to be an essential tool for studying and advancing the field of quantum walks. A crucial feature of Hiperwalk is its enhanced efficiency through the utilization of heterogeneous high-performance computing. In the following section, we outline the central component of the package, which is responsible for attaining the targeted performance level.

\section{Inner Core}\label{sec:4}

Parallel programming, particularly the use of accelerators like modern GPUs, serves as an excellent alternative to enhance performance, particularly for application domains dealing with large-scale problems such as quantum computing simulations. More specifically, quantum walks~\cite{Por18book} can significantly benefit from employing linear algebra routines running on massively parallel hardware to speed up results.

However, writing parallel programs is far from a simple task. Developers must consider numerous factors that influence the final performance of the application, including memory locality, data partitioning, parallel activation libraries, concurrency issues, and processor affinity, among others. When considered collectively, these factors substantially increase the complexity of developing parallel applications. This complexity is further exacerbated when using GPUs, which require a specific programming model and specialized knowledge and training~\cite{OpenCL, CUDAProgModel}. Additionally, none of these aspects are functional with respect to the application being developed, which promotes the coupling of the problem being solved with the hardware resources utilized.

It is important to note that developers who need to build parallel applications often invest considerable time and effort in handling the specific details of low-level hardware. In general, the tools available reinforce this trend, prioritizing performance over providing abstractions for handling details, as discussed in~\cite{TeseNumina}. Another crucial aspect is that as hardware architecture evolves to increase performance, it becomes more sophisticated, increasing its usage complexity and rendering existing software obsolete.

One approach to overcoming these challenges is to employ a layer that abstracts the hardware on which the application executes. As demonstrated in~\cite{7803669}, interpreted programming languages have been successfully used to provide features for parallel applications. A significant advantage of interpreted languages is hardware abstraction, allowing the runtime to transparently evolve alongside hardware updates. This is particularly important when considering the use of GPUs, which evolve rapidly, affecting existing applications.

Another aspect to consider is the provision of linear algebra functions and algorithms for the user. Many libraries are available, such as~\cite{cuBLAS, cuSparse, 10.1145/2925426.2926256, 9092322,10.1145/3204919.3204924, CLSparse, 10.1145/2304576.2304624}. Some libraries are so specialized that they handle only a single aspect of linear algebra, as seen in~\cite{7284398, 9101654, 5493382, 8546266, 6969545, 7095579, 7529927}. Each of these libraries offers different implementations. If application programmers use several of them, they will end up with incoherent code that handles various external APIs, which would impose an unnecessary burden on a research team working with quantum walks.

Quantum walk applications can improve performance by utilizing sparse matrix operations. However, as noted earlier, their implementations are typically separate from standard dense matrix libraries. Moreover, there are not only numerous libraries providing efficient sparse-matrix algorithm implementations but also a multitude of sparse matrix representation formats that can influence the final performance results, as presented in~\cite{10.1145/3017994}.

In light of this situation, researchers requiring parallel implementations of linear algebra functionality should seek easy-to-use, stable, and coherent software tools that manage all the infrastructure needed to accelerate calculations, and provide a homogeneous API to simplify application development.

Neblina~\cite{neblina} is a programming language that relies on OpenCL~\cite{OpenCL} to offer hardware abstraction. Building on this, Neblina provides a subset of linear algebra functions as language constructs with its straightforward implementation, delivering a homogeneous API and shielding users from low-level hardware idiosyncrasies. However, as a complete programming language, it was difficult to evolve Neblina to support new features.

Hiperwalk's inner core is a reimagining of the Neblina programming language as a modular and extensible library. We provide application developers with the benefits of a high-level, homogeneous linear algebra API, while offering system developers a straightforward integration API to support new hardware and other low-level linear algebra implementations. This approach ensures productivity, flexibility, and robustness at both ends of the library, which is impossible with a complete programming language implementation without sacrificing execution performance.

\subsection{Evolutionary barriers}

We outline the main issues that prompted the refactoring of Neblina into a library. One less apparent issue is the tight coupling with OpenCL, which complicates the reuse of other linear algebra libraries and leads to execution errors due to environment configuration. Additionally, supporting hardware without an OpenCL implementation runtime is virtually impossible due to OpenCL-related code dispersed throughout the entire implementation. Even if it were feasible to support new non-OpenCL compliant hardware, we would still need to recompile the whole language system to incorporate the new features.

Another problem is that while Neblina provides an integration API with other programming languages based on C, if the user programs in Python (as in Hiperwalk), creating a C module is less efficient for facilitating inter-program communication. This characteristic is especially important to us since, from the first version of Hiperwalk, integration was achieved by persisting data to the file system and then invoking the Neblina interpreter in a separate process, requiring data to be read all over again. This approach makes it more challenging to integrate and debug the application.

An external factor that also impacted the language is its strong reliance on OpenCL and the availability of hardware runtime implementations. One example of this influence is AMD's decision to discontinue support for CPUs as OpenCL devices\footnote{https://www.amd.com/en/support/kb/release-notes/rn-amdgpu-unified-linux-21-40-1}, limiting execution on machines based on this type of CPU. An alternative is to use an open implementation like Portable OpenCL~\cite{POCL}. However, it does not provide any optimal implementations like vendors can; instead, it uses the system thread library. In fact, this reliance on OpenCL proves our point regarding the effects that the evolution of libraries and hardware have on software lifespan and obsolescence.

\subsection{From language to library}

We now present our motivation for decomposing the Neblina programming language into a shared library and its accompanying Python extension module, which functions as a wrapper around it.

We opted to extract the primary Neblina features and create a shared library defining a homogeneous and centralized API for the subset of linear algebra functions. On top of this, we have a Python extension module that initializes everything. The Hiperwalk frontend relies on this Python extension to access the underlying computational resources, including GPU HPC.

We removed all OpenCL-related code from the main library module and turned the algorithms into more abstract ones, leaving the OpenCL implementation on a separated library that is dynamically loaded during runtime. This is a simpler approach than trying to leverage the OpenCL dependency even with OpenMP related code. Instead, we used OpenCL and OpenMP, transparently, only on separated modules.

This approach significantly simplifies system evolution since we can modify any necessary component and provide only the altered parts. Moreover, if new hardware is released, we can create a library to support it and provide this new software piece independently, without causing side effects for users who do not use the new hardware\footnote{It is noteworthy that we refer to the release of new classic computing hardware that can be used for linear algebra acceleration. Due to our focus on linear algebra, it would not make sense to try to abstract quantum processing units at this library level.}.

Regarding language maintenance, we now rely on Python to offer all the necessary programming language constructs for expressing application programs.

We can summarize the set of benefits as follows:
\begin{enumerate}
  \item Decoupling from OpenCL through an abstract API;
  \item High-level language availability through a Python wrapper that exposes\texttt{final\_time}$=50/\Delta t$ the library functions;
  \item Concentrate efforts on the linear algebra routines;
  \item Increased flexibility to integrate with existing linear algebra libraries through the same API;
  \item Enhanced fine-grained testing capabilities;
  \item Improved performance by retaining data in the application program memory space.
\end{enumerate}

In the subsequent sections, we will delve into the specifics of our development process, providing a more detailed explanation of the steps involved and the decisions made along the way. This will offer a better understanding of the intricacies of our approach and its benefits.

\subsection{Abstract API}

According to~\cite{TeseNumina}, using layers to abstract hardware and separate system-level development from application-level programming offers many benefits. Therefore, we adopt a similar approach to define our new API.

\begin{figure}[!ht]
\centering
  \includegraphics[width=0.4\linewidth]{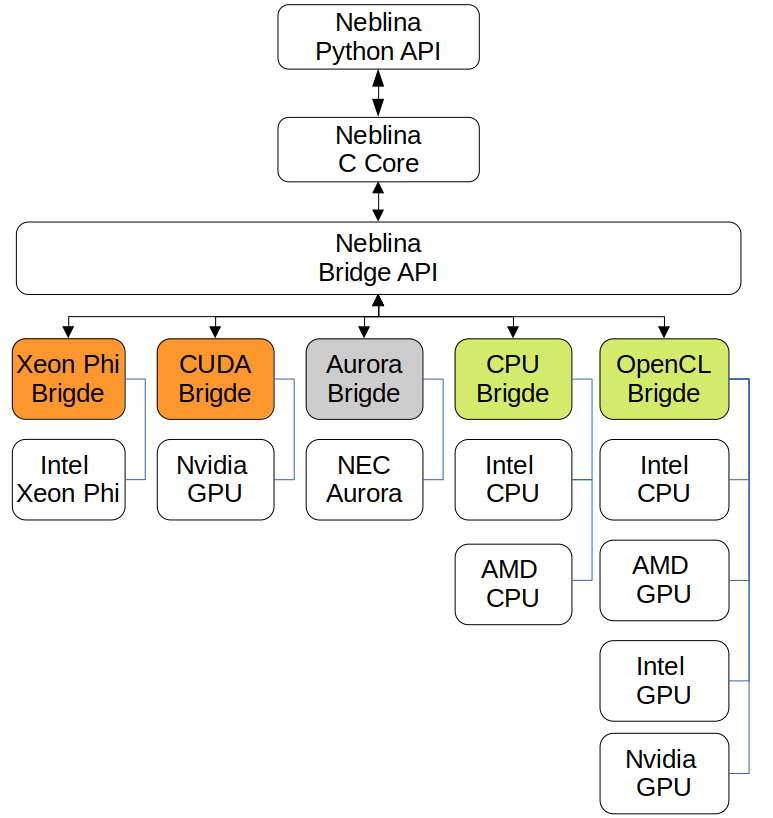}
  \caption{Neblina Core model.}
  \label{fig:neblinaCore}
\end{figure}
In Figure~\ref{fig:neblinaCore}, we can see the layers that comprise the model. At the top layer, we have a Python wrapper that provides the end-user a coherent API to access dense and sparse matrix operations. Next, we have the C implementation of the memory management routines and abstract data structures on the following layer. After that, there is a bridging API that forms the connection between the high-level definition and a specific runtime that requires libraries or implementations dependent on the available accelerator. In green, we see the bridges developed by our team; in gray, a bridge developed by another team for specific hardware; and finally, in orange, the bridges scheduled for development by our team to support native NVidia libraries and Intel Xeon accelerators.

The layers can be detailed as follows:
\begin{itemize}
\item Neblina Python API - this library adheres to the Python definition for an extension module and offers the application programmer a high-level API for creating programs that transparently benefit from low-level parallelism.
\item Neblina C Core - this library implements the necessary handling for loading specific bridges, depending on the hardware available on the machine. It also handles any generic programming required by memory management.
\item Neblina Bridge API - this is part of Neblina Core and defines the set of functions that a low-level implementation (Bridge) must provide to be loaded and executed as a Neblina Core Bridge.
\item Specific Bridge library - this is where we can create specific implementations, and it is responsible for controlling and managing the particular details of each runtime hardware (for example, allocating memory on a GPU). Additionally, we can use this approach to employ different linear algebra implementations like CLBlast\cite{10.1145/3204919.3204924} or experiment with various sparse matrix representations like the ones found in~\cite{10.1145/3017994}. As an example, we moved all the original Neblina code for handling OpenCL devices to our OpenCL Bridge.
\end{itemize}

\subsection{Python wrapper}

We present below a Python program that uses the Neblina core to execute a vector-matrix multiplication using the available GPU.

\begin{figure}
\begin{verbatim}
   1  from neblina import *
   2  def main() 
   3    n = 1000
   4    niter = 1000
   5    M = matrix_new(n,n,float_)
   6    p = vector_new(n,float_)
   7    for i in range(n):
   8        vector_set(p, i, 
                       0.0, 0.0)
   9    move_vector_device(p)
   10   for i in range(n):
   11      for j in range(n):
   12         matrix_set(M, 
                         i, j, 
                         1.0/n, 
                         0.0)
   13   move_matrix_device(M)
   14   vector_set(p, 0, 1.0, 0.0)
   15   for i in range(niter):
   16       p = matvec_mul(p, M)
   17 init_engine(0)
   18 main()
   19 stop_engine()
\end{verbatim}
\caption{A Python program that uses the Neblina API.}
\label{fig:PythonNeblinaCode}
\end{figure}

In Figure~\ref{fig:PythonNeblinaCode}, we can see that in lines 5 and 6, the user declares data structures with their data type explicitly defined, which is similar to the approach used in the prevalent NumPy library. In line 8, we set data using a setter function. Then, in line 9, we request to copy data to the device being used. On line 12, we set data in the matrix with a similar syntax and move data to the device on line 13. What is entirely new are the instructions on lines 17 and 19, where we initialize the Neblina engine to control the device, and after the execution of line 18, we stop the engine and clean all the memory used.

A significant achievement here is maintaining the user data in the same memory space for both the application program and the Neblina library. Of course, if we are dealing with a bridge that controls a device like a GPU, we will have to copy that data to the device's memory at some point. We do this with explicit commands for memory movement; however, we avoid moving data between different operating system processes, which dramatically improves its usage.

\subsection{Flexibility and unit tests}

By separating our library into modules with distinct responsibilities, we achieve a substantial level of flexibility. At the same time, we attain code isolation, which makes it easier to create unit tests for individual functions.

We are currently using our linear algebra routines, initially developed for the Neblina language, as the core implementation of our OpenCL bridge. We also have a pure CPU bridge that employs OpenMP\footnote{https://www.openmp.org/} for code parallelism and can be utilized when no GPU is available on the system.

This level of separation provides excellent opportunities for experimentation with different implementations, particularly when exploring new approaches for sparse matrix operations within GPUs, like the alternatives presented in~\cite{10.1145/3017994}. Now we can implement different versions of the same bridge and experiment with various scenarios to achieve the best possible performance.

Furthermore, the layering provided by our implementation allows us to write tests for each level more independently, significantly contributing to the confidence we can have in our code. For instance, during our alpha development, we were able to identify and correct some memory management issues that would have been much more challenging to find without unit and integration tests.

\subsection{Preliminary results}

We continue to evolve the library while developing new bridges for the Hiperwalk simulator. As a result, we anticipate further developments, and API modifications may be required. Nonetheless, we have already accomplished some relevant results.

\subsubsection{Quantitative results}

Here, we present various evaluations conducted to determine if there were any performance impacts on the library compared to the original Neblina language.

For the performance assessment, we employed a simple vector-matrix multiplication program that executes numerous iterations of the multiplication step (the parallel code). The same system was used, featuring a Core i7 10750H with six cores (12 HyperThreads) and an NVidia GeForce RTX 2060 with 6GB and 1920 CUDA cores. The system operates on the NVidia 495.29.05 proprietary driver and OpenCL 3.0 from CUDA 11.5.56.

In Figure~\ref{fig:PythonTestCode}, we showcase the Python code snippet utilized for the tests. It is quite straightforward; \texttt{n} represents the order of the matrix, while \texttt{it} indicates the number of iterations the program will perform. As this number increases, we may eventually encounter calculations that overflow, resulting in not-a-number values. However, for the purpose of gauging execution time, this is sufficient for our needs.

To record the runtime of the programs, we captured the start and stop times for the entire program and the vector-matrix multiplication loops (which constitute the parallel portion). We measured execution times using the facilities provided by each programming language, but we have omitted this code for readability in the paper.

\begin{figure}
\begin{verbatim}
   1  def test_function():
   2      n = 15000
   3      it = 100000
   4      vec_f = vector_new(n, complex_)
   5      for i in range(n):
   6          vector_set(vec_f, i, 2.0, 2.0)
   7      mat_f = matrix_new(n, n, complex_)
   8      for i in range(n):
   9          for j in range(n):
  10              matrix_set(mat_f, i, j, 
  11                         3.0, 3.0)
  12      move_vector_device(vec_f)
  13      move_matrix_device(mat_f)
  14      res = matvec_mul(vec_f, mat_f)
  15      for i in range(it):
  16          res = matvec_mul(res, mat_f)
\end{verbatim}
\caption{Python code for vector-matrix multiplication.}
\label{fig:PythonTestCode}
\end{figure}

On line 6 of Figure~\ref{fig:PythonTestCode}, we use the \texttt{vector\_set} function to pass the real and imaginary values to be set on the vector, and on line 10, we use \texttt{matrix\_set} with the same concept. Lines 12 and 13 move data to the device; currently, we opt for this to be explicitly requested by the user. Again, the crucial execution takes place from lines 14 to 16, where we call the parallel function inside the for loop.

We also used an equivalent Neblina program and executed each program version five times and calculated the average time value.

\begin{figure}
\centering
  \includegraphics[width=0.7\linewidth]{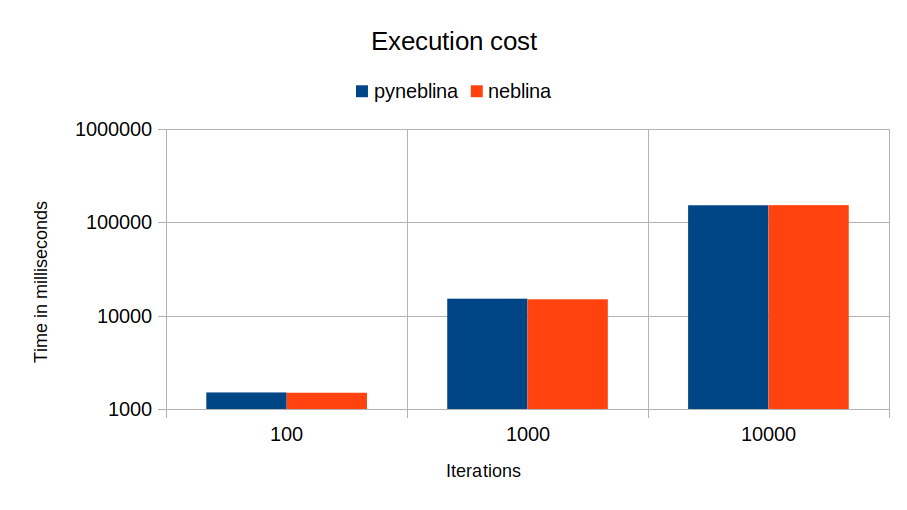}
  \caption{GPU execution times in logarithmic scale.}
  \label{fig:GPUExecutionTime}
\end{figure}

In Figure~\ref{fig:GPUExecutionTime}, we see that when we consider the GPU execution, the times are compatible, meaning that our implementation as a shared library had no side effects on its performance.

\subsubsection{Qualitative results}

When considering adapting code to provide a higher level of abstraction for the user, it is crucial to assess the work qualitatively. In this regard, we identify four significant accomplishments made possible only by the proposed new model. The fact that we achieved these milestones, while still developing the library, reinforces that it was the best decision to pursue.

\begin{enumerate}
\item Creation of new Neblina functions - During the Hiperwalk code modernization, four new functions were required. They were easily added to the library and implemented on the OpenCL Bridge, as there was no need for adaptation to a parser or a syntax analyzer.
\item Support for NEC Aurora\footnote{https://www.nec.com/en/global/solutions/hpc/sx/vector\_engine.html} - Another team implemented the bridge for the NEC Aurora hardware using its own optimized linear algebra libraries. This was made possible due to the separation of responsibilities provided by our layered model. Preliminary execution times can be found in~\cite{223578}.
\item Pure CPU Bridge - We now have a functional CPU implementation, as AMD no longer supports OpenCL for CPUs.
\item Unit tests - As software grows, it becomes increasingly challenging to ensure that all components function as intended. A set of unit tests can significantly improve this aspect. However, with tightly coupled software, it is virtually impossible to accurately assess code quality. In contrast, our separation made it possible to identify and correct some memory management issues.
\end{enumerate}

\section{Conclusions}\label{sec:5}

In conclusion, the new version of the Hiperwalk package offers a comprehensive solution for simulating quantum walks on graphs using heterogeneous high-performance computing. The paper thoroughly outlines the main commands of the package, facilitating users in leveraging its full potential. With its Python frontend and C-based inner core, Hiperwalk seamlessly combines accessibility and computational efficiency. The package's inner core effectively executes parallel linear calculations, including matrix-to-matrix and matrix-vector multiplications, to accommodate large-scale simulations and complex graphs with reduced computational time.

The Hiperwalk package serves as a valuable asset to researchers in the fields of quantum walks and quantum algorithms, providing a robust, user-friendly, and efficient means to explore quantum information processing. Its open-source nature fosters collaboration and innovation within the scientific community, promoting the development of new features and improvements. We encourage researchers to utilize and contribute to Hiperwalk, aiding in the advancement of quantum computing research and the discovery of novel quantum algorithms and spatial quantum search techniques.

Looking ahead, our future work aims to further expand the range of applications for quantum walks within the Hiperwalk package. By enhancing the frontend capabilities, we aspire to accommodate a broader spectrum of quantum walk scenarios and use cases. Simultaneously, we plan to refine the inner core by optimizing load balancing when distributing linear algebra calculations among the processors. 

In the ongoing development of the frontend, we can point the following:
\begin{itemize}
  \item Support for other types of graphs, like bipartite graphs, Johnson Graph, Gridline graph among others.
  \item Support for other types of quantum walks, like staggered and Szegedy.
  \item Users will be able to plot the success probability and optimal running time as a function of the number of vertices. This feature is particularly useful for researchers interested in understanding the performance of quantum walk-based search algorithms and optimizing their design.
\end{itemize}

In the ongoing development of the inner core, we are working on several new features, which include the following:

\begin{itemize}
  \item Utilize different linear algebra routines, including proprietary ones, by creating appropriate runtime bridges to maximize the benefits of each platform's available resources. We are using this approach in our CUDA-based bridge, where we consider the CUDA proprietary libraries for dense and sparse matrix operations (separate libraries in the CUDA ecosystem). This is possible due to our abstraction layers. As a result, we can now manage two different external libraries within a single, compact, and coherent module, with well-defined responsibilities for handling the details of the CUDA libraries. Furthermore, we plan to execute benchmark tests to compare the CUDA bridge against the OpenCL bridge running on top of the CUDA runtime since there could be implementation-specific optimizations that can not be replicated on the OpenCL implementation.
  
  \item Implement matrix tiling to break down computations of instances larger than the available GPU memory. This feature is crucial for supporting more extensive computations since most GPUs have significantly less memory for calculations than the main system.
  
  \item Enable multi-GPU execution within a single machine. We will implement this feature transparently to the Neblina core library, handling the details locally on the OpenCL bridge and the CUDA bridge. These bridges have distinct methods for expressing multi-device computation control and management.

\end{itemize}

By incorporating these new features, we aim to enhance the performance, flexibility, and capabilities of the inner core, further improving its usefulness and efficiency for various applications.

These improvements will lead to even more efficient and versatile simulations, enabling researchers to tackle increasingly complex problems in quantum computing and explore new frontiers in the field of quantum walks.

\section*{Acknowledgment}

The authors acknowledge financial support from CNPq and Faperj.


\begin{thebibliography}{10}

\bibitem{NC00}
M.~A. Nielsen and I.~L. Chuang.
\newblock {\em Quantum computation and quantum information}.
\newblock Cambridge University Press, New York, 2000.

\bibitem{LLP15}
P.~C.~S. Lara, A.~Le{\~a}o, and R.~Portugal.
\newblock Simulation of quantum walks using {HPC}.
\newblock {\em J. Comp. Int. Sci.}, 6:21, 2015.

\bibitem{neblina}
Pedro Lara.
\newblock Neblina, 2015.

\bibitem{AAKV01}
D.~Aharonov, A.~Ambainis, J.~Kempe, and U.~Vazirani.
\newblock Quantum walks on graphs.
\newblock In {\em Proc. 33th STOC}, pages 50--59, New York, 2001. ACM.

\bibitem{FG98}
E.~Farhi and S.~Gutmann.
\newblock Quantum computation and decision trees.
\newblock {\em Phys. Rev. A}, 58:915--928, 1998.

\bibitem{Por18book}
R.~Portugal.
\newblock {\em Quantum Walks and Search Algorithms}.
\newblock Springer, Cham, 2nd edition, 2018.

\bibitem{MP08}
F.~L. Marquezino and R.~Portugal.
\newblock The {QW}alk simulator of quantum walks.
\newblock {\em Comput. Phys. Commun.}, 179(5):359--369, 2008.

\bibitem{BBW11}
S.~D. Berry, P.~Bourke, and J.~B. Wang.
\newblock {QwViz}: {V}isualisation of quantum walks on graphs.
\newblock {\em Comput. Phys. Commun.}, 182(10):2295 -- 2302, 2011.

\bibitem{IW15}
J.~A. Izaac and J.~B. Wang.
\newblock py{CTQW}: A continuous-time quantum walk simulator on distributed
  memory computers.
\newblock {\em Comput. Phys. Commun.}, 186:81 -- 92, 2015.

\bibitem{FRW17}
P.~ E. Falloon, J.~Rodriguez, and J.~ B. Wang.
\newblock {QSW}alk: A {M}athematica package for quantum stochastic walks on
  arbitrary graphs.
\newblock {\em Comput. Phys. Commun.}, 217:162 -- 170, 2017.

\bibitem{10.1007/978-3-319-27119-4_17}
Pei Zhang, Jiabin Yuan, and Xiangwen Lu.
\newblock Quantum computer simulation on multi-gpu incorporating data locality.
\newblock In Guojun Wang, Albert Zomaya, Gregorio Martinez, and Kenli Li,
  editors, {\em Algorithms and Architectures for Parallel Processing}, pages
  241--256, Cham, 2015. Springer International Publishing.

\bibitem{DERAEDT2007121}
K.~{De Raedt}, K.~Michielsen, H.~{De Raedt}, B.~Trieu, G.~Arnold, M.~Richter,
  Th. Lippert, H.~Watanabe, and N.~Ito.
\newblock Massively parallel quantum computer simulator.
\newblock {\em Computer Physics Communications}, 176(2):121--136, 2007.

\bibitem{DERAEDT201947}
Hans {De Raedt}, Fengping Jin, Dennis Willsch, Madita Willsch, Naoki Yoshioka,
  Nobuyasu Ito, Shengjun Yuan, and Kristel Michielsen.
\newblock Massively parallel quantum computer simulator, eleven years later.
\newblock {\em Computer Physics Communications}, 237:47--61, 2019.

\bibitem{Jones2019}
Tyson Jones, Anna Brown, Ian Bush, and Simon~C. Benjamin.
\newblock Quest and high performance simulation of quantum computers.
\newblock {\em Scientific Reports}, 9(1):10736, Jul 2019.

\bibitem{CHEN2018964}
Zhao-Yun Chen, Qi~Zhou, Cheng Xue, Xia Yang, Guang-Can Guo, and Guo-Ping Guo.
\newblock 64-qubit quantum circuit simulation.
\newblock {\em Science Bulletin}, 63(15):964--971, 2018.

\bibitem{Villalonga2019}
Benjamin Villalonga, Sergio Boixo, Bron Nelson, Christopher Henze, Eleanor
  Rieffel, Rupak Biswas, and Salvatore Mandr{\`a}.
\newblock A flexible high-performance simulator for verifying and benchmarking
  quantum circuits implemented on real hardware.
\newblock {\em npj Quantum Information}, 5(1):86, Oct 2019.

\bibitem{Guerreschi2020}
Gian~Giacomo Guerreschi, Justin Hogaboam, Fabio Baruffa, and Nicolas P~D
  Sawaya.
\newblock Intel quantum simulator: a cloud-ready high-performance simulator of
  quantum circuits.
\newblock {\em Quantum Science and Technology}, 5(3):034007, may 2020.

\bibitem{10.1145/2661136.2661156}
Andreas Stefik and Stefan Hanenberg.
\newblock The programming language wars: Questions and responsibilities for the
  programming language community.
\newblock In {\em Proceedings of the 2014 ACM International Symposium on New
  Ideas, New Paradigms, and Reflections on Programming \&amp; Software},
  Onward! 2014, pages 283--299, New York, NY, USA, 2014. Association for
  Computing Machinery.

\bibitem{UCAM-CL-TR-902}
Raoul-Gabriel Urma.
\newblock Programming language evolution.
\newblock Technical Report UCAM-CL-TR-902, University of Cambridge, Computer
  Laboratory, February 2017.

\bibitem{1438333}
Yaofei Chen, R.~Dios, A.~Mili, Lan Wu, and Kefei Wang.
\newblock An empirical study of programming language trends.
\newblock {\em IEEE Software}, 22(3):72--79, 2005.

\bibitem{10.1145/3204919.3204924}
Cedric Nugteren.
\newblock Clblast: A tuned opencl blas library.
\newblock In {\em Proceedings of the International Workshop on OpenCL}, IWOCL
  '18, New York, NY, USA, 2018. Association for Computing Machinery.

\bibitem{Klockner:2012:PPS:2109228.2109321}
Andreas Kl{\"o}ckner, Nicolas Pinto, Yunsup Lee, Bryan Catanzaro, Paul Ivanov,
  and Ahmed Fasih.
\newblock Pycuda and pyopencl: A scripting-based approach to gpu run-time code
  generation.
\newblock {\em Parallel Comput.}, 38(3):157--174, March 2012.

\bibitem{10.1145/3419111.3421287}
Vaishaal Shankar, Karl Krauth, Kailas Vodrahalli, Qifan Pu, Benjamin Recht, Ion
  Stoica, Jonathan Ragan-Kelley, Eric Jonas, and Shivaram Venkataraman.
\newblock Serverless linear algebra.
\newblock In {\em Proceedings of the 11th ACM Symposium on Cloud Computing},
  SoCC '20, pages 281--295, New York, NY, USA, 2020. Association for Computing
  Machinery.

\bibitem{CG04}
A.~M. Childs and J.~Goldstone.
\newblock Spatial search by quantum walk.
\newblock {\em Phys. Rev. A}, 70:022314, 2004.

\bibitem{OpenCL}
Khronos Group.
\newblock Opencl, 2010.

\bibitem{CUDAProgModel}
Developer NVidia~Zone.
\newblock Cuda programming model, 2016.

\bibitem{TeseNumina}
Paulo Rogerio~da Motta.
\newblock {\em Abstra{\c c}{\~a}o para Programa{\c c}{\~a}o Paralela: Suporte
  para o Desenvolvimento de Aplica{\c c}{\~o}es}.
\newblock PhD thesis, Departamento de Inform{\'a}tica, March 2012.

\bibitem{7803669}
R.~Ribeiro and P.~Motta.
\newblock Towards a gpu abstraction for lua.
\newblock In {\em 2016 International Symposium on Computer Architecture and
  High Performance Computing Workshops (SBAC-PADW)}, pages 13--18, Oct 2016.

\bibitem{cuBLAS}
Developer NVidia~Zone.
\newblock cublas, 2016.

\bibitem{cuSparse}
Developer NVidia~Zone.
\newblock cusparse, 2016.

\bibitem{10.1145/2925426.2926256}
Linnan Wang, Wei Wu, Zenglin Xu, Jianxiong Xiao, and Yi~Yang.
\newblock Blasx: A high performance level-3 blas library for heterogeneous
  multi-gpu computing.
\newblock In {\em Proceedings of the 2016 International Conference on
  Supercomputing}, ICS '16, New York, NY, USA, 2016. Association for Computing
  Machinery.

\bibitem{9092322}
T.~{Gautier} and J.~V.~F. {Lima}.
\newblock Xkblas: a high performance implementation of blas-3 kernels on
  multi-gpu server.
\newblock In {\em 2020 28th Euromicro International Conference on Parallel,
  Distributed and Network-Based Processing (PDP)}, pages 1--8, March 2020.

\bibitem{CLSparse}
Clmathlibraries.
\newblock Clsparse, 2015.

\bibitem{10.1145/2304576.2304624}
Bor-Yiing Su and Kurt Keutzer.
\newblock Clspmv: A cross-platform opencl spmv framework on gpus.
\newblock In {\em Proceedings of the 26th ACM International Conference on
  Supercomputing}, ICS '12, pages 353--364, New York, NY, USA, 2012.
  Association for Computing Machinery.

\bibitem{7284398}
C.~{Yang}, Y.~{Wang}, and J.~D. {Owens}.
\newblock Fast sparse matrix and sparse vector multiplication algorithm on the
  gpu.
\newblock In {\em 2015 IEEE International Parallel and Distributed Processing
  Symposium Workshop}, pages 841--847, May 2015.

\bibitem{9101654}
J.~{Lee}, S.~{Kang}, Y.~{Yu}, Y.~{Jo}, S.~{Kim}, and Y.~{Park}.
\newblock Optimization of gpu-based sparse matrix multiplication for large
  sparse networks.
\newblock In {\em 2020 IEEE 36th International Conference on Data Engineering
  (ICDE)}, pages 925--936, April 2020.

\bibitem{5493382}
A.~H. {El Zein} and A.~P. {Rendell}.
\newblock From sparse matrix to optimal gpu cuda sparse matrix vector product
  implementation.
\newblock In {\em 2010 10th IEEE/ACM International Conference on Cluster, Cloud
  and Grid Computing}, pages 808--813, May 2010.

\bibitem{8546266}
N.~v.~{Kondratyev}, M.~G. {Persova}, Y.~G. {Soloveichik}, and D.~S. {Kiselev}.
\newblock Using hyb sparse matrix storage format for solving linear systems
  obtained by fem discretization on gpu.
\newblock In {\em 2018 XIV International Scientific-Technical Conference on
  Actual Problems of Electronics Instrument Engineering (APEIE)}, pages
  135--139, Oct 2018.

\bibitem{6969545}
B.~{Neelima}, G.~R.~M. {Reddy}, and P.~S. {Raghavendra}.
\newblock Predicting an optimal sparse matrix format for spmv computation on
  gpu.
\newblock In {\em 2014 IEEE International Parallel Distributed Processing
  Symposium Workshops}, pages 1427--1436, May 2014.

\bibitem{7095579}
K.~{He}, S.~X.~. {Tan}, H.~{Wang}, and G.~{Shi}.
\newblock Gpu-accelerated parallel sparse lu factorization method for fast
  circuit analysis.
\newblock {\em IEEE Transactions on Very Large Scale Integration (VLSI)
  Systems}, 24(3):1140--1150, March 2016.

\bibitem{7529927}
R.~{Eberhardt} and M.~{Hoemmen}.
\newblock Optimization of block sparse matrix-vector multiplication on
  shared-memory parallel architectures.
\newblock In {\em 2016 IEEE International Parallel and Distributed Processing
  Symposium Workshops (IPDPSW)}, pages 663--672, May 2016.

\bibitem{10.1145/3017994}
Salvatore Filippone, Valeria Cardellini, Davide Barbieri, and Alessandro
  Fanfarillo.
\newblock Sparse matrix-vector multiplication on gpgpus.
\newblock {\em ACM Trans. Math. Softw.}, 43(4), jan 2017.

\bibitem{POCL}
Pekka J{\"a}{\"a}skel{\"a}inen, Carlos~S{\'a}nchez de~La~Lama, Erik Schnetter,
  Kalle Raiskila, Jarmo Takala, and Heikki Berg.
\newblock pocl: A performance-portable opencl implementation.
\newblock {\em International Journal of Parallel Programming}, 43(5):752--785,
  2015.

\bibitem{223578}
F{\'e}lix D.~P. Michels, Philippe O.~A. Navaux, Paulo Motta, and Renato
  Portugal.
\newblock Simulating quantum walks on vector processors.
\newblock In {\em Proc. of 2nd WQuantum}, pages 25--30, may 2022.

\end{thebibliography}

\end{document}